PALAEOBOTANY

A tree without leaves

Brigitte Meyer-Berthaud and Anne-Laure Decombeix

*The puzzle presented by the famous stumps of Gilboa, New York, finds a solution in the discovery of two fossil specimens that allow the entire structure of these early trees to be reconstructed.*

The Middle Devonian (397–385 million years ago) was a notable time for the evolution of early land plants: the diversification of reproductive strategies, the advent of leaf-precursors, and a tendency to increasing height all led to the rise of plants of modern appearance. One of the best locations for studying these processes is Gilboa, New York, which is internationally renowned for its wide range of terrestrial organisms from the late Middle Devonian. These organisms include a rich variety of arthropods, for example, but also large, enigmatic stumps that in life formed extensive stands and were fossilized where they grew.

Since the discovery of these *Eospermatopteris* stumps in the nineteenth century, questions about their affinities and what kind of trees they supported have exercised generations of palaeobotanists. Evidence presented by Stein *et al.*[1] on page 000 of this issue not only answers these questions but also provides a new glimpse of the requirements necessary to make a tree.

Reconstructing entire fossil plants is an important step in assessing the patterns of plant diversification through time, and the roles that plants played in past environments. This task is challenging because plants naturally shed parts of their body during their lifetime, and it is rarely achieved for structures such as trees that have a complex architecture and extended lifespan. However, the two exceptionally well-preserved specimens collected by Stein *et al.*[1] were sufficient to reconstruct the kind of tree that made the Gilboa forest (Fig. 1a).

The two specimens partly overlap in size and morphology. One consists of a slender trunk exceeding 6 m in length, exhibiting an enlarged *Eospermatopteris*-type base, and covered with branch scars at the top. Slender 'coalified' strands extending downwards from the bottom of the stump are interpreted as roots. The other specimen is the top part of a trunk terminated by a crown of short and erect branches, and with branch scars at the bottom. Branches of the crown were divided digitately — that is, they produced several closely spaced branchlets flattened in a plane. The terminal appendages were three-dimensional and do not resemble leaves (organs with a flat blade of photosynthetic tissue). In the Devonian, this branching pattern characterizes the Pseudosporochnales, an extinct group traditionally considered to be closer to the ferns than to the seed plants and known to have included 3 meter tall organisms about 392 million years ago[2]. This group was successful and occurred worldwide during the Middle Devonian[3].

One consequence of Stein and colleagues' discovery is that *Archaeopteris*, a close relative of seed plants that flourished in the Late Devonian (385–359 million years ago), is no longer the oldest known modern tree (Fig. 1b). The conifer-like reconstruction of this 'progymnosperm' appears in most botany textbooks and many of us are familiar with its woody stump measuring up to 1.5 m in diameter, which is suspected to have reached 40 m in height and was deeply anchored into the soil by an extensive root system[4,5]. *Archaeopteris* had large, perennial (long-lived) branches that were essential components of its architecture[6]. As it grew, *Archaeopteris* added more of these branches to its trunk, a process that contributed to its three-dimensional increase in size. Short-lived branches produced at the periphery were leafy.

The Gilboa tree differs from *Archaeopteris* in having a trunk of more moderate size, in lacking perennial branches, and in possessing a limited root system that apparently consisted of many roots of similar size. The short, disposable branches of the crown were shed when senescent and were continuously replaced by new branches at the top of the trunk. Unlike *Archaeopteris*, the root and branch systems in the Gilboa tree did not increase much in extent during growth. In this, the tree resembles the tree-ferns, and the cycads and palms with single trunks that occur today; but unlike these, it was unable to make planate leaves. Instead it seems that photosynthesis was carried out by the periodically shed branches of the crown and, more specifically, by their three-dimensional terminal appendages.

One benefit often assumed for taller plants is their enhanced ability to capture light. Ten years ago, Niklas[7] simulated the architecture of early land plants and tested their efficiency in performing several essential functions. The Gilboa tree fits closely with the morphology that optimizes two functions, mechanical stability and reproduction. But the reduced surface area of its crown was not optimal for light interception.

Two contrasting ways of making trees evolved during the Devonian (Figs 1a, b; Fig. 2). That represented by *Archaeopteris*, and by most extant trees of temperate and tropical areas, requires a complex machinery of tissues and organs to achieve growth in all spatial directions and build the larger body sizes recorded in the plant kingdom. The Gilboa tree represents an economical alternative where, beyond the necessary investment in spores to ensure reproduction, the products of photosynthesis were mainly devoted to vertical growth of the trunk. The new specimens from New York[1] show that the first giants in the history of the land plants achieved the tree habit and significant biomass despite their inability to construct optimal photosynthetic structures, such as leaves or horizontal branches, and despite not building an extensive root system.

Brigitte Meyer-Berthaud (CNRS) and Anne-Laure Decombeix (Université Montpellier 2) are in the Unité Mixte de Recherche Botanique et Bioinformatique de l'Architecture des Plantes (AMAP), CIRAD, 34398 Montpellier Cedex 5, France.
e-mails: meyerberthaud@cirad.fr; anne-laure.decombeix@cirad.fr

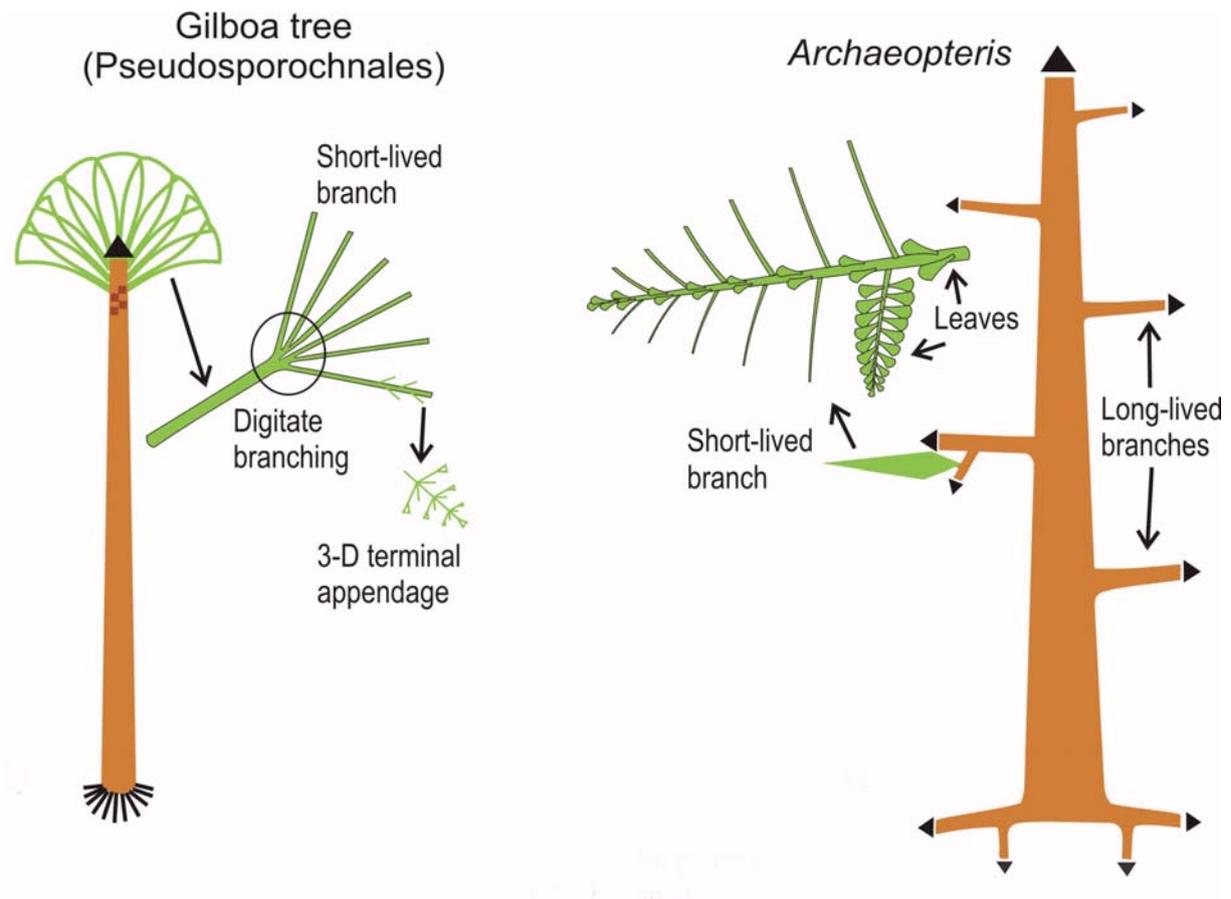

**Figure 1 Two types of Devonian tree.** a, The newly described Gilboa tree[1], a member of the Pseudosporochnales (Fig. 2), had no leaves and a limited root system, and displayed an economical strategy whereby a single long-lived organ, the trunk, grew vertically. b, By contrast *Archaeopteris* possessed leafy twigs, and had long-lived roots and branches that grew together with the trunk. Photosynthetic organs are shown in green; black triangles indicate long-lived organs.

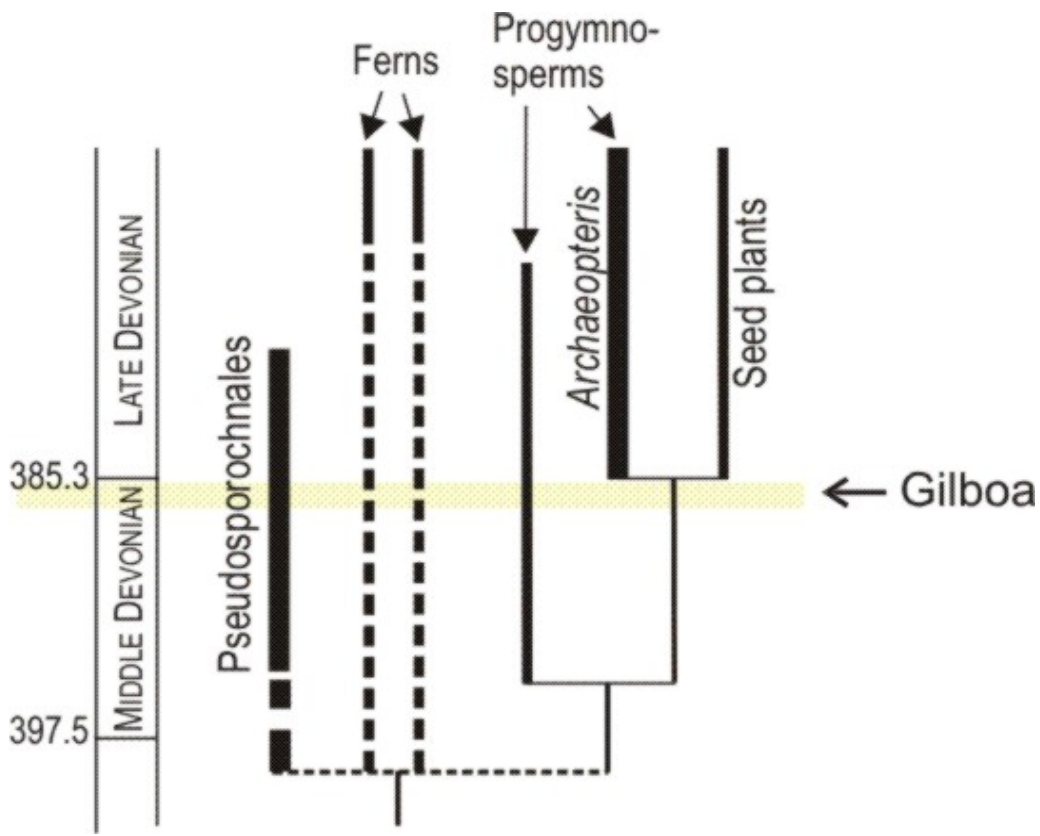

**Figure 2 Trees in time.** Two contrasting ways of making trees, evident in the fossils of the Gilboa tree and *Archaeopteris*, evolved in the Devonian but are still successful today. The Gilboa tree is a member of the extinct group, Pseudosporochnales. *Archaeopteris* is a progymnosperm, a close relative of the seed plants[6,8] that is also extinct. Time scale is millions of years ago. Many representatives of ferns and seed plants exist today, the latter being by far the main constituent of the world's current terrestrial flora.